\documentclass[aps,twocolumn,groupedaddress,pra,floatfix,showpacs]{revtex4}
\pdfoutput=1
\usepackage{graphicx}
\usepackage{tabularx}
\usepackage{dcolumn}
\usepackage{amsmath}
\usepackage{amssymb}
\usepackage{bm}
\usepackage{graphics}
\usepackage{subfigure}

\begin{document}

\newcommand{\bibpath}{E:/MyDocuments/Projecten/H3Plus/H3Plus_Articles/Bibliography/}
\newcommand{\arhplus}{\ensuremath{\mathrm ArH^+}}
\newcommand{\htwo}{\ensuremath{\mathrm H_2}}
\newcommand{\htwoplus}{\ensuremath{\mathrm H_2^+}}
\newcommand{\hhh}{\ensuremath{\mathrm H_3^+}}
\newcommand{\otp}{ortho-to-para ratio}
\newcommand{\pcc}{cm\ensuremath{^{-3}}}
\newcommand{\figwidth}{3.4in}

\title{Resonant structure of low-energy \hhh\ dissociative recombination}

\author{Annemieke Petrignani}
\author{Simon Altevogt}
\author{Max H. Berg}
\author{Dennis Bing}
\author{Manfred Grieser}
\author{Jens Hoffmann}
\author{Brandon Jordon-Thaden}
\author{Claude Krantz}
\author{Mario B. Mendes}
\author{Old\v{r}ich Novotn\'{y}}
\altaffiliation{Present address: Columbia University, 550 West 120th Street, New York, NY 10027, USA}
\author{Steffen Novotny}
\altaffiliation{Present address: Aalto University, Department of Micro and Nanoscicences, FI-00076 Aalto, Finland}
\author{Dmitry A. Orlov}
\author{Roland Repnow}
\author{Tobias Sorg}
\author{Julia St\"{u}tzel}
\author{Andreas Wolf}
\affiliation{Max-Planck-Institut f\"{u}r Kernphysik, Saupfercheckweg 1, D-69117 Heidelberg, Germany}
\author{Henrik Buhr}
\altaffiliation{Present address: Physikalisch-Technische Bundesanstalt, Bun\-des\-al\-lee 100, 38116 Braunschweig, Germany}
\affiliation{Department of Particle Physics, Weizmann Institute of Science, 76100 Rehovot, Israel}
\author{Holger Kreckel}
\altaffiliation{Present address: Department of Chemistry and Astronomy, University of Illinois at Urbana-Champaign, Urbana, IL 61801, USA}
\affiliation{Columbia University, 550 West 120th Street, New York, NY 10027, USA}
\author{Viatcheslav Kokoouline}
\affiliation{Department of Physics, University of Central Florida, Orlando, Florida 32816, USA}
\affiliation{Laboratoire Aim\'e Cotton, CNRS, Universit\'e Paris-Sud XI, 91405 Orsay, France}
\author{Chris H. Greene}
\affiliation{Department of Physics and JILA, University of Colorado, Boulder, Colorado 80309-0440, USA}

\date{\today}

\begin{abstract}
New high-resolution dissociative recombination rate coefficients of rotationally cool and hot \hhh\ in the vibrational ground state have been measured with a 22-pole trap setup and a Penning ion source, respectively, at the ion storage ring TSR. The experimental results are compared with theoretical calculations to explore the dependence of the rate coefficient on ion temperature and to study the contributions of different symmetries to probe the rich predicted resonance spectrum. The break-up energy was investigated by fragment imaging to derive internal temperatures of the stored parent ions under differing experimental conditions. A systematic experimental assessment of heating effects is performed which, together with a survey of other recent storage-ring data, suggests that the present rotationally cool rate-coefficient measurement was performed at 380$^{+50}_{-130}$\,K and that this is the lowest rotational temperature so far realized in storage-ring rate-coefficient measurements on \hhh. This partially supports the theoretical suggestion that higher temperatures than assumed in earlier experiments are the main cause for the large gap between the experimental and theoretical rate coefficients. For the rotationally hot rate-coefficient measurement a temperature of below 3250\,K is derived. From these higher-temperature results it is found that increasing the rotational ion temperature in the calculations cannot fully close the gap between the theoretical and experimental rate coefficients.
\end{abstract}

\maketitle

%_________________________________________________
\section{\label{sec:intro} Introduction}
The rate coefficient of the dissociative recombination (DR) process in \hhh\ has been subject to debate for decades \cite{larsson:2008, larssonorel:2008}. Over the past few years, there has been a convergence in the DR rate coefficient as determined by theoretical calculations and measured in different storage-ring and afterglow experiments. The source for the remaining descrepancy between the reported theoretical and experimental rate coefficients has been opted to be the internal-state distribution of the \hhh\ ions. It has been stated that the \hhh\ rate coefficient is now sufficiently well-known for astrophysical models treating the \hhh\ abundance \cite{larsson:2008}. However, on other astrophysical topics, such as the reigning \hhh\ ortho-para spin temperatures \cite{tom:2009} and the observation of the (3,3) metastable state \cite{goto:2002}, rotational state-specific rate coefficients are needed. Theory and experiment still disagree when it comes to collision-energy and rotational-temperature dependence. Very little agreement exists in the resonance structure of the \hhh\ DR rate coefficient, i.e., in the occurence, amplitudes and positions of resonances. Assignment of the individual resonances in the rate coefficient of this elementary system has until now remained far out of reach. Published storage-ring measurements of the rate coefficient for rovibrationally cold \hhh\ report that only the lowest rotational states of the vibrational ground state are believed to be populated \cite{mccall:2004,kreckel:2005a,petrignani:2009}. Predictions of cold rovibrational \hhh\ rate coefficients differ locally by up to an order of magnitude \cite{santos:2007}.  A recent theoretical investigation \cite{santos:2007} pointed out that better agreement is achieved with the experimental \hhh\ data if a rotational ion temperature of 1000\,K is assumed. The reason for using a higher temperature in Ref. \cite{santos:2007} was to reflect the possibility that higher rotational ion temperatures might have been present than had been assumed appropriate for the experimental conditions. The cold experimental \hhh\ beams could possibly undergo heating during extraction, acceleration, and/or circulation in the storage ring. No evidence for large heating effects inside the ring that could explain a 1000\,K or higher temperature has been observed to date \cite{wolf:2006}. Recently, it has been suggested that, in the case of ion production with the supersonic ion source, the ion-source extraction could have led to heating \cite{kreckel:2010}. Further investigations are necessary to shed light on the discrepancies remaining between theory and experiment in the DR of \hhh.

In fact, in the long history of DR studies, it has not yet been possible to achieve complete quantitative agreement between theory and experiment for {\it any} molecular target ion. The closest to achieving full agreement is probably the diatomic hydrogen ion, but even that system has not yet exhibited the close convergence of DR theory and experiment that has been demonstrated for other observables such as photoabsorption. The emergence during the past decade of a promising theoretical candidate for both the mechanism \cite{KokooulineNature:2001} and the quantitative description of H$_3^+$ DR \cite{kokoouline:2003a,kokoouline:2003b,santos:2007,JungenPratt:2009} has apparently resolved the long-standing uncertainty about the overall low energy recombination rate, but the limitations of the theory remain unclear. By making a detailed resonance-by-resonance test, starting with the present article, this will hopefully elucidate both the strengths and limitations of the theory. It may also prove useful in suggesting systematic issues that could be affecting experimental measurements, such as the ionic rovibronic state distribution and external field effects.

We have measured high-resolution DR rate coefficients for rotationally cool and hot \hhh\ in their vibrational ground state for collision energies between 0 and 30 eV. The rotationally cool \hhh\ beam was produced with a buffer-gas cooled 22-pole trap, while the rotationally hot \hhh\ beam was produced with a Penning ion source. The use of a cold photocathode electron gun provided us with the high resolution \cite{orlov:2004}, and the greatly improved transmission of the low-current \hhh\ beam from the 22-pole trap to the storage ring enabled us to gather more statistics over Ref.\cite{petrignani:2009}. The rotational ion-beam temperatures are investigated through imaging measurements under the conditions of the rate-coeficient measurements. We also present a re-evaluation of a previous measurement with the 22-pole trap \cite{petrignani:2009}, which gives an estimate of the ion beam temperature produced by the 22-pole trap, shedding light on a possible heating mechanism. The dependence of the rate coefficient on ion temperature is investigated through theoretical predictions, and the theory-experiment comparison permits a first attempt to assign the \hhh\ DR resonance structure.

%_________________________________________________
\section{\label{sec:exp} Experiment}

\begin{center}
\begin{table*}
{\small
\caption[]{\label{table:ImExps} The imaging measurements at near 0-eV energy that are presented and discussed in this paper. The rotational temperatures are derived from a $\chi^2$ fit between simulation and measurement with uncertainties given at a 90\% confidence interval. The statistical and systematic errors can add up to an uncertainty of about 135\,K in the rotational ion temperature. The measurements are grouped for each reference. The labels IR refer to the imaging experiments conducted under conditions similar to the rate measurements.}
\renewcommand{\arraystretch}{1.2}
\begin{tabular}{|l|c|c|c|c|c|c|c|c|c|}
\hline 
label [Ref]                     & ion source   & T$_{rot} \pm$ & \multicolumn{4}{c|}{source settings} & \multicolumn{2}{c|}{cooler} 
                                                                                                         							 			 & target \\
                                &              & $(\Delta T)_{stat}$
                                														 & $He$      & $H_2$     & trapping time  
                                                                                         		 & parent	 & kT$_{\perp}$ 
                                                                                         							        & time at 0 eV & time at 0 eV \\
                                &              & (K)  			 & (\pcc)    & (\pcc)    & (ms)  & gas  	 &(meV) & (s)     		 & (s)    \\
\hline
\hline
IR22p                           & 22-pole trap &  $380\pm35$ & $10^{16}$ & $10^{9}$  & 1     & p-\htwo & 12   & 0-8.5  			 & 0-8.5  \\
\hline 
I22p-09 \cite{petrignani:2009}  & 22-pole trap &  $200\pm75$ & $10^{15}$ & $10^{10}$ & 100   & p-\htwo & 13.5 & 0-0.5  			 & 0-7.5  \\
I22n-09 \cite{petrignani:2009}  & 22-pole trap &  $150\pm50$ & $10^{15}$ & $10^{10}$ & 100   & n-\htwo & 13.5 & 0-0.5 			 & 0-7.5  \\
Ipenn-09 \cite{petrignani:2009} & Penning      & $3250\pm35$ &           &           &       & n-\htwo & 13.5 & 0-0.5  			 & 0-7.5  \\
\hline 
Isonic-10 \cite{kreckel:2010}   & supersonic   & $950\pm100$ &           &           &       & p-\htwo & 12.5 & off    			 & 0-10   \\
IRsonic-10 \cite{kreckel:2010}  & supersonic   & $450\pm100$ &           &           &       & p-\htwo & 12.5 & 0-10   			 & 0-10   \\
\hline
\end{tabular}}
\end{table*}
\end{center}

The experimental studies were carried out at the storage ring TSR at the Max-Planck-Institut f\"{u}r Kernphysik in Heidelberg \cite{habs:1989,sprenger:2004}. For the rotationally cool \hhh\ measurements, the ions were produced in a radiofrequency (rf) storage ion source and buffer-gas cooled for 1 ms in a cryogenic 22-pole ion trap at a helium density of $10^{16}$ \pcc\ and a temperature of $\sim$15\,K. It has been shown through spectroscopy measurements and modeling \cite{petrignani:2009,asvany:2009} that the cooling of the \hhh\ in the 22-pole trap is efficient and produces ions with rotational and translational temperatures that agree well with the nominal trap temperature. At 15\,K, the lowest two rotational states, $(N^+,K^+)=(1,1)$ and $(1,0)$, separated by $\Delta=22.8$ cm$^{-1}$, of the vibrational ground state are populated, with $N^+$ the total angular momentum of the ion and $K^+$ the body-frame projection of the angular momentum of the ion. The (1,1) and (1,0) states belong to para and ortho nuclear-spin symmetries, respectively. A differential pumping stage between the rf ion source and the rf ion trap reduces the partial \htwo\ pressure in the trap chamber to 10$^{-8}$ mbar and minimizes any conversions between the ortho and para symmetries. The two states are therefore believed to be equally populated instead of Boltzmann distributed. For each injection into the ring, up to $10^6$ \hhh\ ions were stored in the trap, buffer-gas cooled, and extracted. The rovibrationally hot \hhh\ beam was produced in a Penning ion source.

After extraction, the \hhh\ beam is accelerated to $4.113 \pm 0.002$ MeV and injected into the ring. Transmission from the ion trap to the storage ring has been increased to above 50\% (as opposed to the $<10$\% transmission in Ref. \cite{petrignani:2009}). The transmission of the Penning beam is less critical since high currents ($\mu$A) can be produced. When present, the vibrational excitation has time to decay inside the ring, leaving rotational excitation only after 2 s storage time \cite{kreckel:2002}. In two straight sections of the ring, named electron cooler and electron target, cold electron beams were merged with the \hhh\ ion beam. The function of the electron cooler is to phase-space cool the ion beam. Here, the electron and ion beams are velocity matched to give 0-eV collision energy at an energy spread of 12 meV and an electron density of $\sim10^7$ \pcc\ for the entire storage time of 8.5 s. The electron target can also phase-space cool the ion beam when set to cooling (near 0-eV) energy, but with less efficiency since it has a lower electron density of 10$^6$ \pcc. Primarily, the electron target functions as the reaction region to study the DR reaction. It provides an ultracold electron beam, produced by a nitrogen-cooled photocathode \cite{orlov:2004}, with transverse and longitudinal temperatures of $k$T$_{\perp}\approx$ 1 meV and $k$T$_{\parallel}\approx$ 30 $\mu$eV, respectively. Two types of measurements are described below: rate-coefficient and imaging measurements. Because also a re-evaluation of and comparisons to earlier measurements are carried out here, each performed under different experimental conditions, an overview of the imaging measurements and the rate measurements is given in Tables \ref{table:ImExps} and \ref{table:RateExps}, respectively. The respective measurements will subsequently be referred to using the labels shown in these tables.

\begin{center}
\begin{table}[t]
{\small
\caption[]{\label{table:RateExps} The rate-coefficient measurements presented and discussed in this paper. For all rate measurements, the electron cooler was set to cooling (near 0-eV) energy for the entire storage time. The measurements are grouped for each reference. The 22-pole trap pressures of Ref. \cite{kreckel:2005a} are similar to Ref. \cite{petrignani:2009}. Ref \cite{kreckel:2005a} used trapping times of 1-100 ms and Ref. \cite{petrignani:2009} of 1 ms only.} 
\renewcommand{\arraystretch}{1.2}
\begin{tabular}{|l|c|c|c|c|c|}
\hline
label [Ref]                     & ion          & parent  & \multicolumn{2}{c|}{cooler}  & target  \\
                                & source       & gas     & kT$_{\perp}$       & at 0 eV & at 0 eV \\
                                &              &         & (meV)  					  & (s)     & (s)     \\
\hline
\hline
R22n                            & 22-pole trap & n-\htwo & 12                 & 0-8.5   & 0-2     \\
Rpenn		                        & Penning      & n-\htwo & 12                 & 0-8.5   & 0-2     \\
\hline
R22n-09 \cite{petrignani:2009}  & 22-pole trap & n-\htwo & 13.5               & 0-11.5  & 0-2     \\
Rpenn-09 \cite{petrignani:2009} & Penning      & n-\htwo & 13.5               & 0-11.5  & 0-2     \\
\hline
R22n-05 \cite{kreckel:2005a}    & 22-pole trap & n-\htwo & 13.5               & 0-11.5  & 0-2     \\
\hline
Rsonic-10 \cite{kreckel:2010}   & supersonic   & n-\htwo & 12.5               & 0-10    & 0-2     \\
\hline
\end{tabular}}
\end{table}
\end{center}

The internal temperature of the rotationally cool \hhh\ beam has been investigated by analyzing the dissociation dynamics of \hhh\ breaking up into three H atoms through an imaging experiment at 0($\pm0.5$)-meV collision energy (see IR22p in Table \ref{table:ImExps}). This measurement was performed with both the electron cooler and the electron target at near 0-eV energy for the entire storage time. Events with three H product atoms were recorded with a position sensitive detection system as decribed in Ref. \cite{strasser:2001}. No new imaging measurements were performed with the highly excited beam. Previous imaging data from the Penning source, Ipenn-09, were used instead to estimate an ion temperature for the rotationally hot ion beam.

An improved forward simulation of the dissociations was utilized to derive the effective rotational temperatures from the imaging data. The improvements contained the following main modifications. The anisotropy of the fragment momentum-distribution is now accounted for using the Dalitz plot describing the reaction dynamics presented in Ref. \cite{petrignani:2009}. Additionally, the kinetic energy release has been recalculated using the latest findings on the ionization energy of \hhh\ (see Sec. \ref{sec:res}). And finally, an effective target length approximating the effect of the merging and de-merging of the electron beam on top of the straight overlap between the ion and electron beams has been included. An important difference between the current IR22p measurement and the measurements from Ref. \cite{petrignani:2009} is that the electron cooler was at cooling energy for, respectively, the entire storage time and 0.5 s only. To investigate the effect of the electron cooler, the measurements from Ref. \cite{petrignani:2009} are re-evaluated using the improved simulation.

The rate measurement, R22n, was performed with the electron cooler at cooling conditions for the entire storage time, similar to the imaging measurement, IR22p. This is typical for all rate measurements (see Table \ref{table:RateExps}) as the electron beam of the target is not permanently set to cooling energy as is the case for a 0-eV imaging measurement. The electron beam of the target was set to cooling for the first 2 s only, after which the electron velocity was detuned to investigate the DR rate as function of collison (detuning) energy. The current rate measurements cover the range $E_d=0-30$ eV. The count rate of the neutral fragments produced by the DR reactions in the target was measured with a large surface barrier detector (SBD) with an area of 10x10 cm$^2$ situated at $\sim12$ m downstream from the center of the electron target. Simultaneously, the charged \htwoplus\ products as well as the neutral H or \htwo\ products from collisions of \hhh\ with residual gas in the ring were recorded with a scintillator and the SBD, respectively, as a measure of the ion beam intensity. The general detection and normalisation procedure is described in detail in Ref. \cite{kreckel:2010}. Here, the reference signal was recorded at a collision energy of 10.8 eV. No absolute measurements were performed.

%_________________________________________________
\section{Theory}

The general theoretical approach used to calculate the DR rate coefficient as function of electron collision energy was previously described in detail in Refs. \cite{ kokoouline:2003b,kokoouline04a,kokoouline05b,santos:2007}. An intermediate step in the calculation is the following sum:
 
\begin{eqnarray}
 \frac{1}{2N^+ +1}\sum_N (2N+1)P_{o'}^{N,\Gamma}\,,
\end{eqnarray}

\noindent where $P_{o'}^{N,\Gamma}$ is the probability of dissociative recombination for a single $e^-+$H$_3^+$ collisional event. Here the system of the ion and the incident electron are described by the initial channel ${o'}$ which describes both the initial rovibrational state of the ion $|{o'}\rangle$ and its angular momentum coupling with the incident electron, assumed to be in a $p$-wave throughout. (See for instance, Eq. (4) and the corresponding discussion of Ref. \cite{santos:2007}.) The label $\Gamma$ refers to the symmetry of the initial state of the system (the ion plus the electron at infinity), while $N^+$ and $N$ are the angular momenta of the ion and neutral H$_3^++e^-$ system, respectively. The quantum numbers $\Gamma$ and $N$ are conserved during the collision, along with the total energy $E$.  

The key ingredient in the calculation of the DR probability $P_{o'}^{N,\Gamma}$ is the energy-dependent scattering matrix $S_{o{o'}}^{\rm phys}(E)$, 
representing the scattering of the electron from the initial channel $|{o'}\rangle$ to a final channel $|{o}\rangle$. The scattering matrix  $S^{\rm phys}(E)$ depends on the total energy $E$ of the system, with the electron kinetic energy in channel ${o'}$ equal to $\epsilon_{o'}=E-E_{o'}$ where $E_{o'}$ is the energy of the target ion in channel $o'$. In our theoretical treatment \cite{kokoouline03b,kokoouline04a,kokoouline05b,santos:2007}, the complete energy-normalized wave function of the H$_3^+ +e^-$ system at large distances $r$ between the ion and the incident electron is written in the following way (see also Eq. (2.55) of Ref. \cite{aymar96})

\begin{widetext}
\begin{eqnarray}
\label{eq:wf}
 \psi_{o'}={\cal A} r^{-1}\sum_c \Phi_c(\omega) W_c(r,\nu_c)Z_{c{o'}}(E)+\\
{\cal A} r^{-1}\sum_o \frac{i\Phi_o(\omega)}{\sqrt{2\pi k_o} }\left[ \exp(-i k_or-i\phi_o)\delta_{o{o'}}-\exp(i k_or+i\phi_o)S^{\rm phys}_{o{o'}}(E)\right]\,.\nonumber
\end{eqnarray}
\end{widetext}

\noindent The second sum above represents the scattering part of the wave function: The incident electron arrives in channel $|{o'}\rangle$ and can escape into one of the channels open for ionization. Those energetically-open ionization channels are labeled with an index $o$. In the above expression, we have explicitly written the electronic wave function of the incident electron only ($r$-dependence). The two other electrons of H$_3^+$ are considered to be frozen, but accounted for implicitly, because the scattering matrix (or quantum defect function used to calculate $S^{\rm phys}(E)$) is obtained with those electrons present in the system, of course. The factor $\Phi_o(\omega)$ represents the ionic rovibrational wave function with $\omega$ the spatial and spin coordinates of the ion as well as the angular and spin coordinates of the scattering/Rydberg electron, i.e., including the angular and spin wavefunctions of the outermost electron and corresponding angular momentum coupling factors. The first sum in Eq. (\ref{eq:wf}) describes the part of the wave function in channels closed for ionization. The factor $\Phi_c(\omega)$ represents the corresponding rovibrational wave function of the ionic core, again with the angular and spin degrees of freedom of the outermost electron. $W_c(r,\nu_c)$ is the exponentially decaying electronic radial function in the closed channel, unity normalized, and ${\cal A}$ denotes an antisymmetrization operator for the electron. The coefficient $Z_{c{o'}}(E)$ gives the relative contribution (amplitude) of the given closed channel to the total wave function: The relative probability to find the system in closed channel $|c\rangle$  is $|Z_{c{o'}}(E)|^2$, and this can be used to classify resonances. In a standard scattering theory formulation, an additional term would also be present to describe  the dissociation channels of the neutral molecular system. In our treatment, use of a complex absorbing potential in the vibrational degrees of freedom permits us to describe the loss of dissociative flux without explicitly computing individual scattering matrix elements into the dissociation channels. Lastly, in Eq. (\ref{eq:wf}), the phase factor $\phi_o$ denotes the usual Coulomb phase contributions, i.e., $\phi_o = -l_o\,\pi/2 + \sigma_o + (1/k_o)\,ln\,2k_o r$. 

Like the energy-dependent scattering matrix $S^{\rm phys}(E)$, the coefficients $Z_{c{o'}}(E)$ are calculated from the smooth scattering matrix $S_{i'i}$ \cite{aymar96}, approximated here as energy-independent over an energy range 1-2 eV. First, the matrix

\begin{eqnarray}
\label{eq:tildeZ}
 \tilde Z_{c{o'}}(E)= e^{-i\pi\hat\nu_c} \left( S_{cc}-e^{-2i\pi\hat\nu_c}\right)^{-1} S_{c{o'}} \,,
\end{eqnarray}

\noindent is calculated. Here, $\hat\nu_c$ is the  $N_c\times N_c$ diagonal matrix  with nonzero elements $\nu_c=1/\sqrt{2(E_c-E)}$ (effective quantum numbers and atomic units); $E_c$ is the ion energy for the closed channel $|c\rangle$. The matrix  $\tilde Z(E)$ corresponds to energy-normalized electronic wave functions in closed channels. Therefore, to obtain amplitudes $Z(E)$ for unity-normalized electronic functions, which correspond more closely to standard spectroscopy classifications, we have to multiply each element of $\tilde Z(E)$ with $\nu_c^{3/2}$:

\begin{eqnarray}
\label{eq:Z}
Z_{c{o'}}(E)=\nu_c^{-3/2} \tilde Z_{c{o'}}(E)\,.
\end{eqnarray}

\noindent Eqs. (\ref{eq:tildeZ}) and (\ref{eq:Z}) give the relative probability $|Z_{c{o'}}(E)|^2$, proportional to the fraction of time spent by the electron in the closed channel $c$ when incident in channel $o'$, at any energy $E$. Because every channel (closed or open) is characterized with a specific ionization threshold energy $E_c$ or $E_o$, the effective quantum number $\nu_d$ for the dominant closed channel $|d\rangle$ is also obtained immediately. For energies $E$ far from a resonance, several closed channels may have competing contributions. At a resonance, one closed channel usually dominates clearly, although for a ``complex resonance'' multiple closed channels can be important for DR \cite{curik:2007}.

%_________________________________________________
\section{\label{sec:res} Results and Discussion}

\subsection{Imaging Results}

\begin{figure}
{\includegraphics[width=\figwidth]{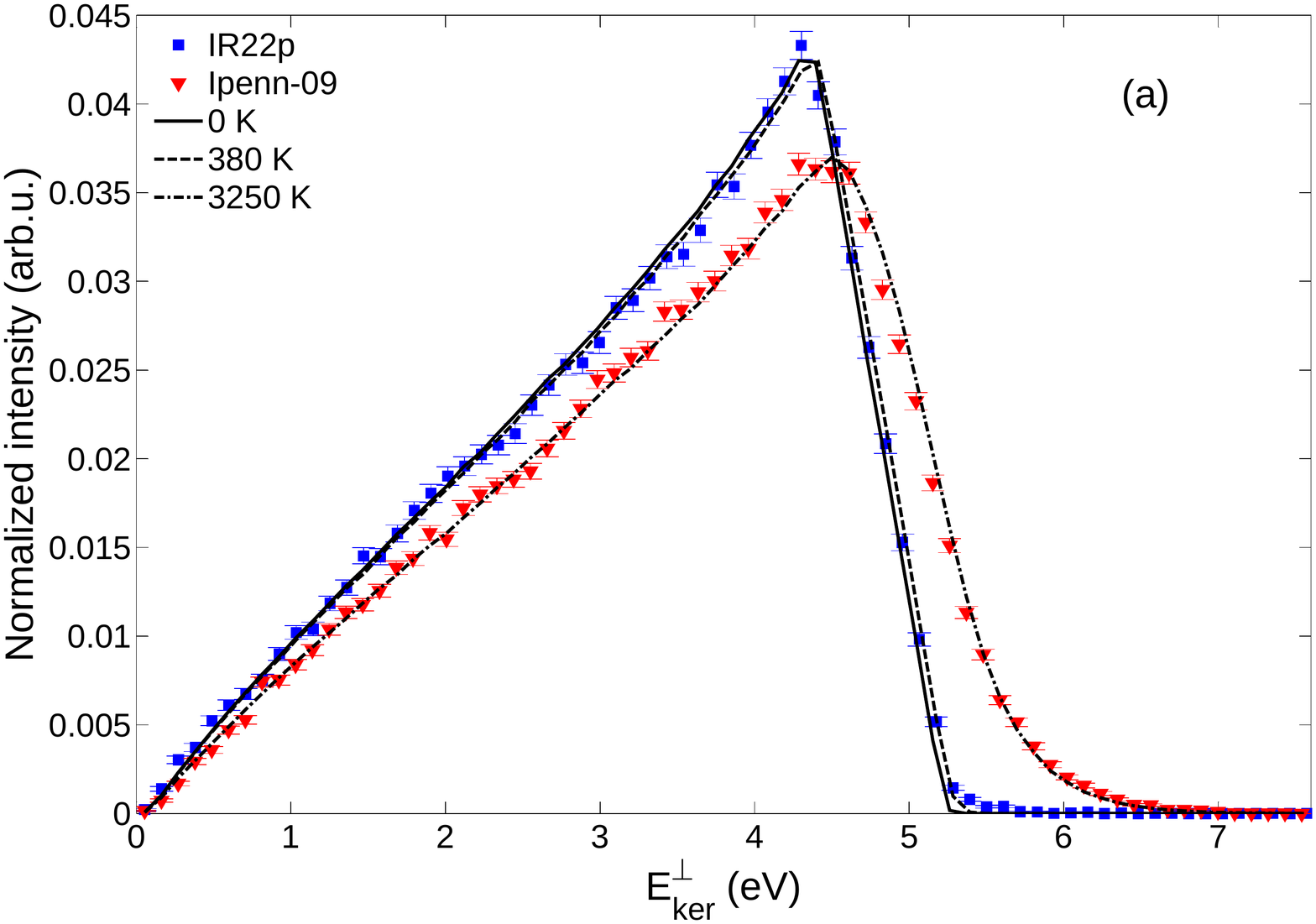}}
\\
{\includegraphics[width=\figwidth]{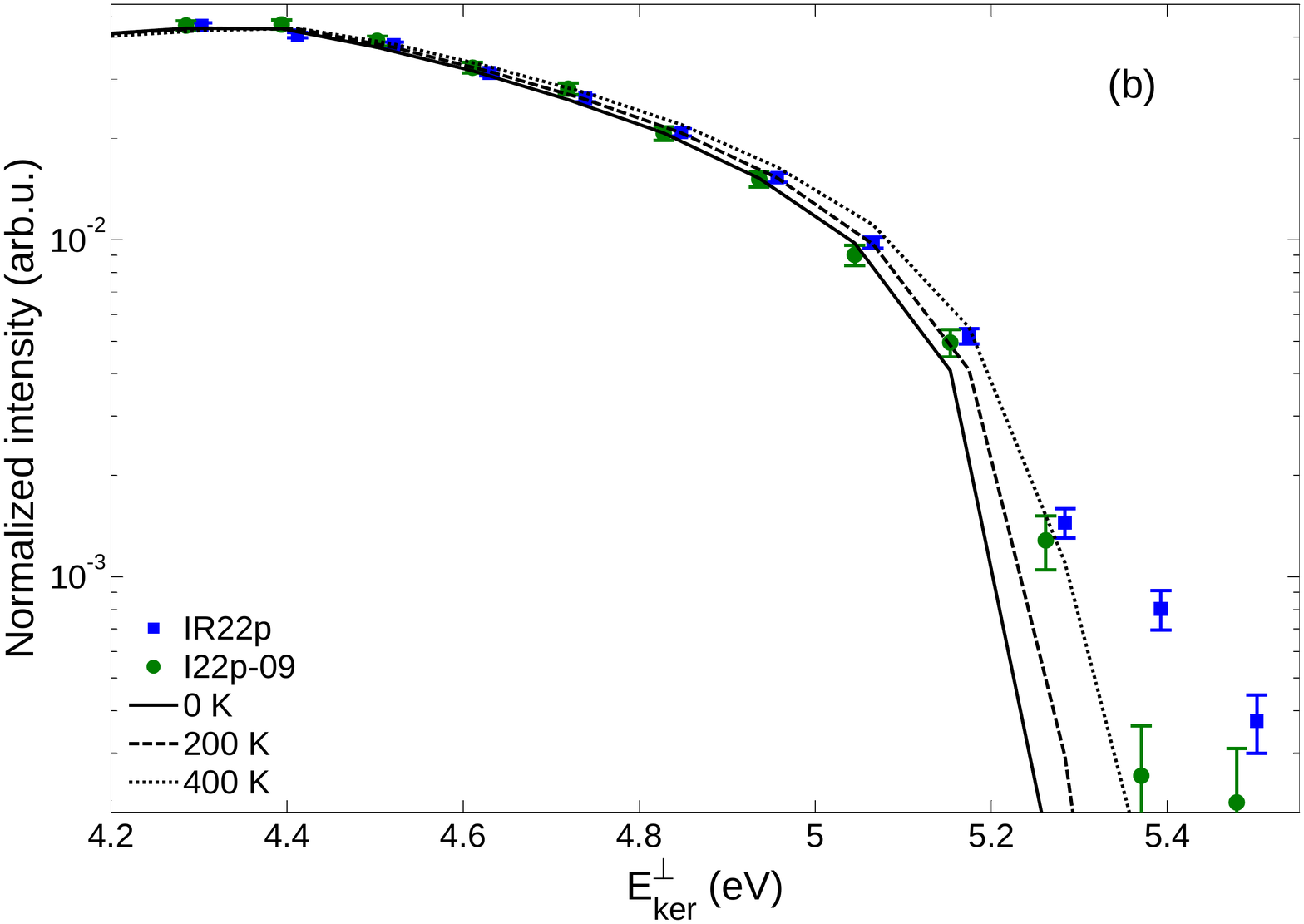}}
\caption{(a) The measured 2D distributions of the kinetic energy released in the DR of \hhh\ into H + H + H. In blue, the distribution for the \hhh\ beam produced with para-\htwo\ in the 22-pole trap, IR22p, and in red, the distribution for the \hhh\ beam produced with the Penning source, Ipenn-09. Also shown are the simulated distributions at 0, 380, and 3250\,K. All distributions are normalized to unit area. (b) The semilogarithmic zoom of the tail of IR22p (blue) and the previously measured I22p-09 (green). The simulations shown are at temperatures of 0, 200, and 400\,K.}
\label{fig:Eker}
\end{figure}

The 2D kinetic energy distributions of IR22p and Ipenn-09 are plotted in Fig. \ref{fig:Eker}a). These distributions are projections of the total energy released, $E^\perp_{ker}$, to the three H atoms upon dissociation, thereby revealing information on the amount of internal energy available in the parent ion. The temperatures of the stored ion beams are derived with forward Monte-Carlo simulations, which describe the DR experiment from dissociation up to detection and analyze the simulated events like real data. The Dalitz plot presented in Ref. \cite{petrignani:2009} is used to include the linear geometrical bias of the dissociation dynamics. An effective electron target-length is included to account for the overlap and merging and de-merging regions between the electron and ion beams. Using Eq. (6) of Ref. \cite{kreckel:2010}, in both the measured and simulated data, the distance $s\sim12$ m from the center of the electron target to the detector is applied to give the tranversal energy $E^\perp_{ker}$ in terms of the inter-particle distances $R$. As the rotational population and the state-dependent DR rate coefficients are unknown, a Boltzmann distribution is modeled assuming a rotationally independent rate. The rotational temperatures are derived from a $\chi^2$ fit between the simulation and the measurement using the temperature as free parameter. A temperature of 380\,K could be determined for IR22p (see also Table \ref{table:ImExps}). A re-evaluation of Ipenn-09 gave a roughly ten times higher temperature of 3250\,K, in good agreement with the initially reported $\sim3500$\,K \cite{petrignani:2009}. As can be observed in Fig. \ref{fig:Eker}a), the simulations and measurements deviate slightly, which is probably due to the approximations made. The shape of the simulated distribution, however, comes much closer to the measured data then before, when the linear geometrical bias from the Dalitz data was not included.

The relative uncertainty, estimated from a 90\% confidence interval of the $\chi^2$ fit, is about 35\,K. Typically, the minimized $\chi^2$ was around 3. The systematic uncertainty is larger with $\sim135$\,K, dominantly determined by the uncertainty in the target length. In fact, the target length is estimated to be around 1.10 m, which is closer to the minimum than to the maximum overlap possible, giving rise to an asymmetric error bar ($\sim_{-130}^{+50}$\,K). The systematic errors coming from the uncertainties of the ion-beam energy ($\sim30$\,K) and kinetic energy release ($\sim30$\,K for 7 meV difference) have been significantly reduced as described below.

The improved value for the \hhh\ dissociation energy \cite{kutzelnigg:2006,galster:2001,helm:2010} has led to an improved determination of the kinetic energy release in the \hhh\ DR. This kinetic energy release has been recalculated to be 4.790 eV (from $N^+=1,K^+=1$) using the theoretical \hhh\ dissociation energy that over the years has converged to 4.330 eV \cite{kutzelnigg:2006}. The experimental \hhh\ dissociation energy of 4.337 eV \cite{galster:2001,helm:2010}, previously reported to be 4.373(21) eV \cite{cosby:1988}, gives only a 7 meV difference with theory, which yields about 30\,K difference in the ion temperature. Additionally, the ion-beam energy for the measurements reported in Ref. \cite{petrignani:2009} has now been analyzed in more detail, giving a more precise value of $4.096 \pm 0.002$ MeV.

Figure \ref{fig:Eker}b) shows the tail of the 2D kinetic energy distribution IR22p together with the previously measured imaging data of the DR of rotationally cool \hhh, I22p-09. As can be seen, the distribution I22p-09 is colder than IR22p and even approaches the displayed 0-K simulation between 4.8 and 5 eV, whereas IR22p follows the 200-K curve more closely in this region. Above $\sim5.1$ eV, where the amplitude has dropped more than an order of magnitude already, the distributions approach the hotter curves, indicative of a non-Boltzmann character. Using the improved simulation and the updated energies to determine the temperatures of I22p-09 (shown) and I22n-09 (not shown), the minimum reduced $\chi^2$ with 90\% confidence interval gives $200 \pm 75$\,K and $150 \pm 50$\,K, respectively. Together with the systematic uncertainties of $\sim135$\,K, the 22-pole ion-source temperature of $\sim15$\,K cannot be entirely excluded. The re-evaluated temperatures are in agreement with the previous analysis stating $<$300\,K. The use of para-\htwo\ or normal-\htwo\ to produce the \hhh\ has no significant effect on the ion temperature of the stored beam.

The main difference between IR22p and I22p-09 is the use of the electron cooler. The temperatures derived for IR22p ($\sim380$\,K) and I22p-09 ($<300$\,K) are associated to the electron cooler on cooling energy for the entire storage time and 0.5 s only, respectively. Both measurements also differ in trapping times and gas pressures (see Table \ref{table:ImExps}), however, these are not expected to yield different temperatures. The lower (higher) He pressure is compensated for by the longer (shorter) trapping time and the trapping times should attain sufficient cooling power at the used buffer-gas densities as reported in Ref. \cite{kreckel:2005a}. This points to the electron cooler as source of heating for the ion beam of the 22-pole trap. In Ref. \cite{kreckel:2010}, at higher temperatures, the opposite effect is observed, where the electron cooler cools the ion-beam temperature of the supersonic ion source from 950\,K to 450\,K (see Table \ref{table:ImExps}). The heating of the 22-pole beam as opposed to the cooling of the supersonic beam by the electron cooler is suggestive of the electron cooler reaching an equilibrium between 380 and 450\,K. The tranverse energy spread of the electrons in the cooler of $\sim12$ meV might be a factor in preventing cooling to lower temperatures. The temperature of the ion beam produced by the supersonic ion source is best approximated with a temperature of 950\,K when the electron cooler is not used. Mutatis mutandis, the \hhh\ ion temperature after extraction from the 22-pole trap adheres to the $<300$\,K temperature and it is inconclusive whether any heating already occured in this case. All rate coefficients, however, have been measured with the electron cooler turned on for the entire storage time. The rate coefficients measured with the 22-pole trap will therefore correspond to the heated temperature of $\sim380$\,K. Similarly, the rate coefficient measured with the supersonic ion source relates to the cooled temperature of $\sim450$\,K.

No imaging data exist for the rotationally hot \hhh\ beam from the Penning source with the electron cooler used for the entire storage time. Assuming the electron cooler is cooling at this high temperature, an ion-beam temperature below 3250\,K is believed to correspond to the DR rate coefficient of the highly excited \hhh\ beam.

\subsection{Rate Coefficient Measurements}

\begin{figure}
\centering
{\includegraphics[width=\figwidth]{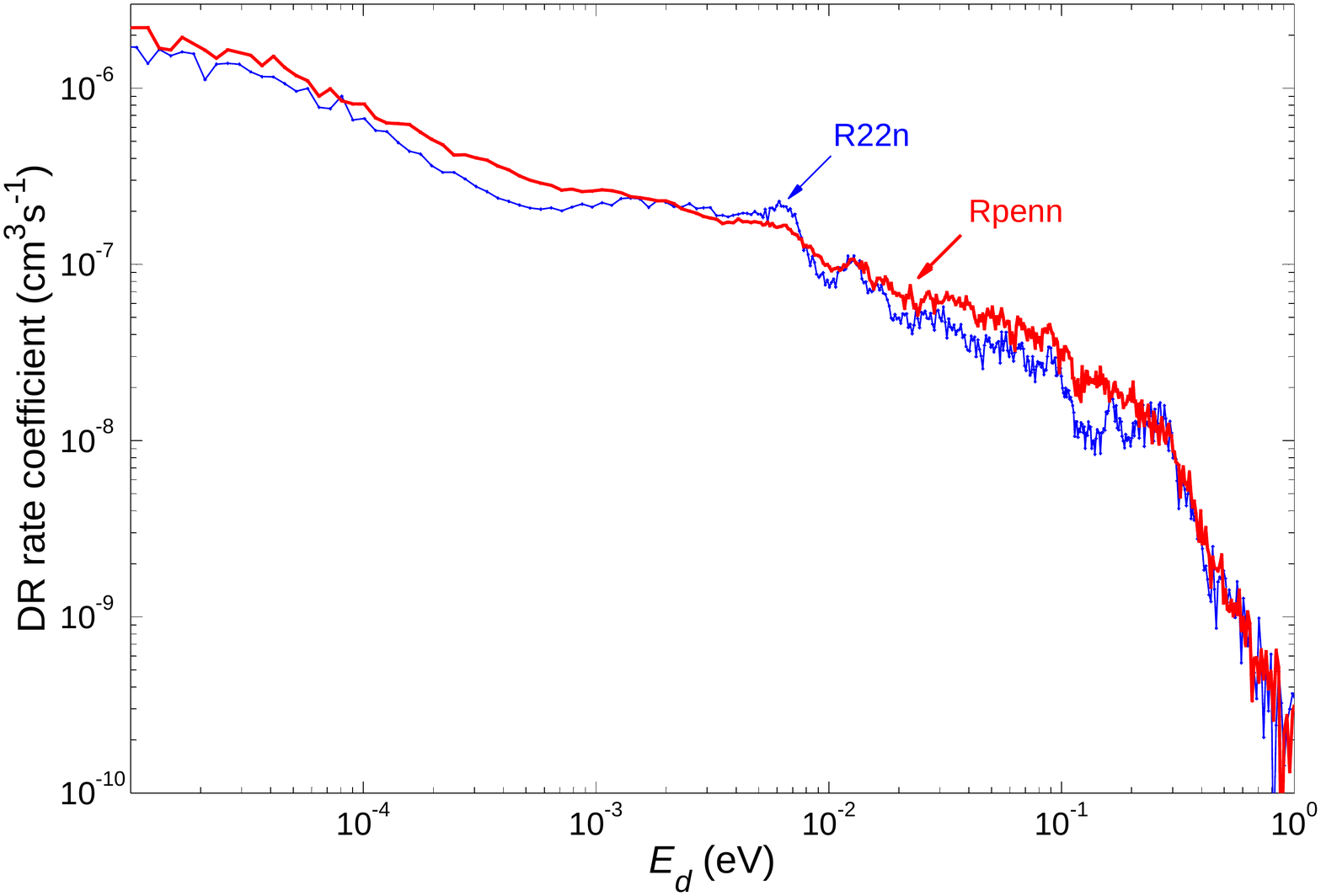}}
{\caption{\label{fig:ExpRateslog}\ The \hhh\ DR rate coefficients, R22n and Rpenn, as measured for the ion beam produced with the 22-pole trap and the Penning source, respectively, on a double-logarithmic scale.}}
\end{figure}

Figure \ref{fig:ExpRateslog} shows the high-resolution rate coefficients measured with high statistics as functions of the collision energy on a double logarithmic scale, before the reduction performed in the detailed comparisons below. Except for the energy region of 2-8 meV, the rotationally hot \hhh\ rate coefficient Rpenn is larger than or equal to the rotationally cool rate R22n. Distinct structures are observed for the colder beam with the structures in particular below 20 meV being more clearly resolved than ever before. The rate coefficient for highly excited ions is the highest resolution measurement for excited \hhh\ to date. Although much smoother, this rate coefficient still shows a remnant of the cold rate coefficient structures.

\begin{figure}
\centering
{\includegraphics[width=\figwidth]{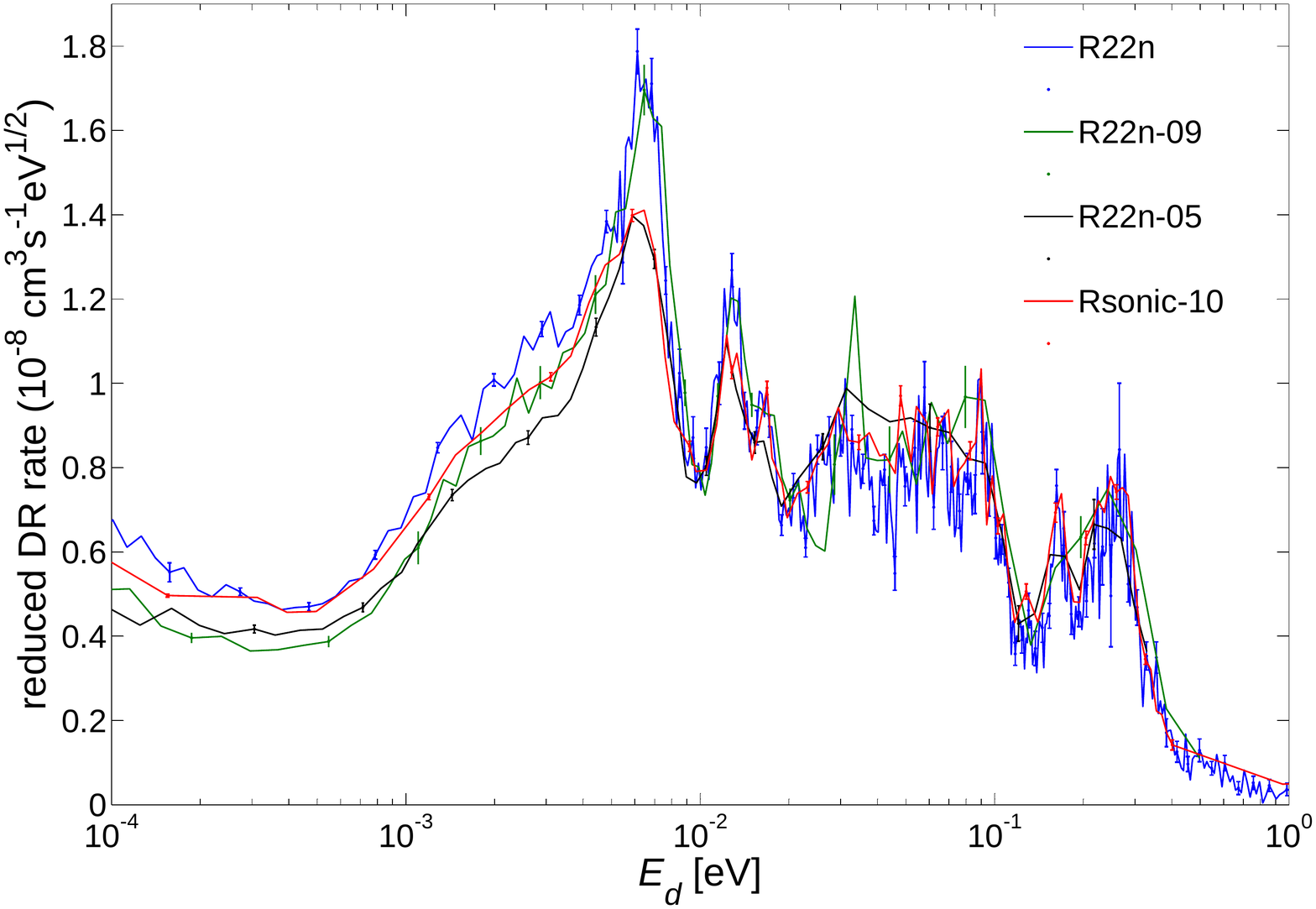}}
{\caption{\label{fig:compareExps}\ The reduced DR rate coefficients from the rotationally cool \hhh\ beam R22n compared with previously reported rates on a semi-logarithmic scale. For clarity, error bars are only shown for every 5th point.}}
\end{figure}

The rate coefficients of the current and previous measurements (see Table \ref{table:RateExps}) with rotationally cool \hhh\ are plotted in Fig. \ref{fig:compareExps}. Here, the reduced DR rate coefficients are shown, which are the product $\langle\alpha\rangle\sqrt{E_{d}}$ and serve to eliminate the $1/\sqrt{E_{d}}$ threshold dependence of the rate coefficient $\alpha$ at low collision energies. All rate coefficients, except R22n-09, are normalized to their 10-eV peak (not shown) with the absolute scale taken from Ref. \cite{mccall:2003b} or Ref. \cite{kreckel:2010} in the case of Rsonic. R22n-09 has no 10-eV peak to normalize to. Previously, it was normalized to the 12-meV peak under the assumption that this resonance is temperature independent. As can be observed in Fig. \ref{fig:compareExps} this may not be the case and the R22n-09 rate coefficient is instead normalized to the minimum around 10 meV of R22n. Assuming the 10-meV normalization to be correct, the rate coefficients R22n and R22n-09 look similar up to at least 20 meV. Above, comparison is difficult due to the large statistical errors of R22n-09. Both rates are expected to be similar as they have been measured under similar conditions. However, an even older measurement of the rate coefficient using the 22-pole trap as injector, R22n-05, exhibits less pronounced resonances around 6 and 12 meV, similar to the rate coefficient Rsonic from the supersonic ion source. The cause for the lack of structure in this region is presently not understood. It is conceivable that the lower amplitudes of the 6- and 12-meV resonances is the result of hotter temperatures, similar to the resonance behaviour observed in the theoretical calculations discussed below. The ion temperature of Rsonic was estimated to lie around 450\,K under the conditions of a rate measurement. This is slightly higher than the 380\,K estimated to match R22n and R22n-09.

\subsection{Rate Coefficient Predictions}

\begin{figure}
\centering
{\includegraphics[width=\figwidth]{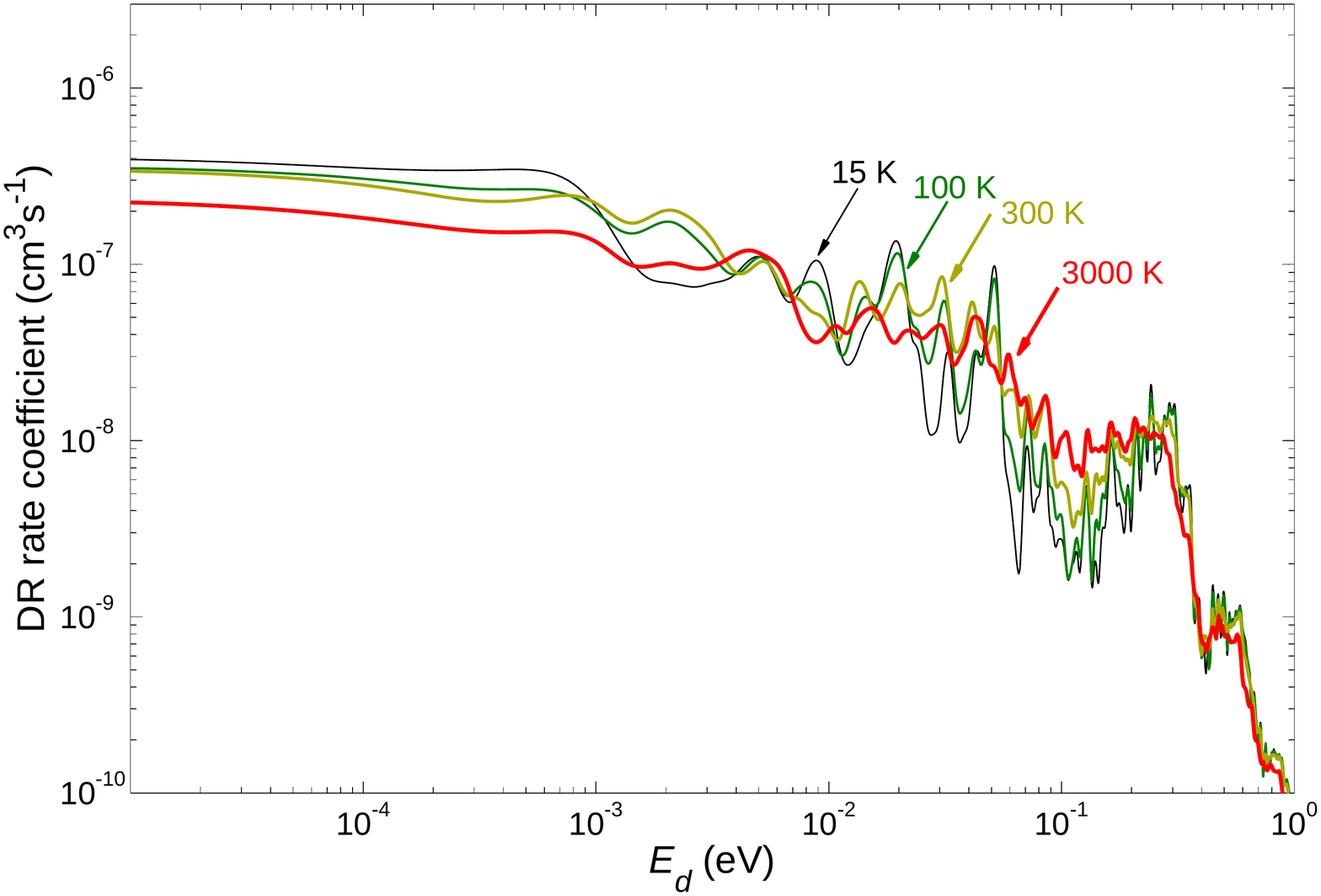}}
{\caption{\label{fig:TheorRatelog}\ The \hhh\ DR rate coefficients as predicted for 15, 100, 300, and 3000\,K, with the line thickness increasing with temperature. The 15\,K rate is calculated using a 1:1 \otp.}}
\end{figure}

Figure \ref{fig:TheorRatelog} displays theoretical rate-coefficients adapted from Ref. \cite{santos:2007} for temperatures of 15, 100, 300, and 3000\,K convolved with energy resolutions of $\Delta E_{\shortparallel}=25\,\mu$eV and $\Delta E_\perp=500\,\mu$eV, similar to experiment. The experimental values given in Sec. \ref{sec:exp} are somewhat more conservative, however, we do not expect this minor difference to be of importance. The relative para-/ortho-ratio of \hhh\ ions (including the multiplicity factor) used are in thermal equilibrium, $n_p/n_o=\exp(\Delta/k_BT)/2$, except for the 15\,K rate coefficient that assumes equal para and ortho populations (see Sec. \ref{sec:exp}). Using equal or thermalized populations for higher temperatures makes little difference. The predictions show a generally higher rate for the hot \hhh\ than for the cold \hhh, but with some local differences, as observed in the experiments. The rate-coefficient increase is most pronounced between 20 and 200 meV, where the rate increases monotonically with increasing temperature. Above 200 meV, the rate coefficient decreases monotonically with increasing temperature. Three resonances around 9, 20, and 60 meV emerge with decreasing temperature, in between which there is no monotonic behaviour. It is expected that as the ion temperature decreases, the thermal averaging increases the contribution from the lower angular momentum states, and any individual resonance seen at low temperature should appear more strongly than in a higher temperature spectrum. Furthermore, the change in the amplitude of the rate coefficient seems to decrease with increasing temperature. Whereas the step from 15 to 300\,K changes the rate considerably, a generally smaller effect is observed when increasing the temperature from 300 to 3000\,K.

\subsection{Study of the Resonant Structure}

\begin{figure}
\centering
{\includegraphics[width=\figwidth]{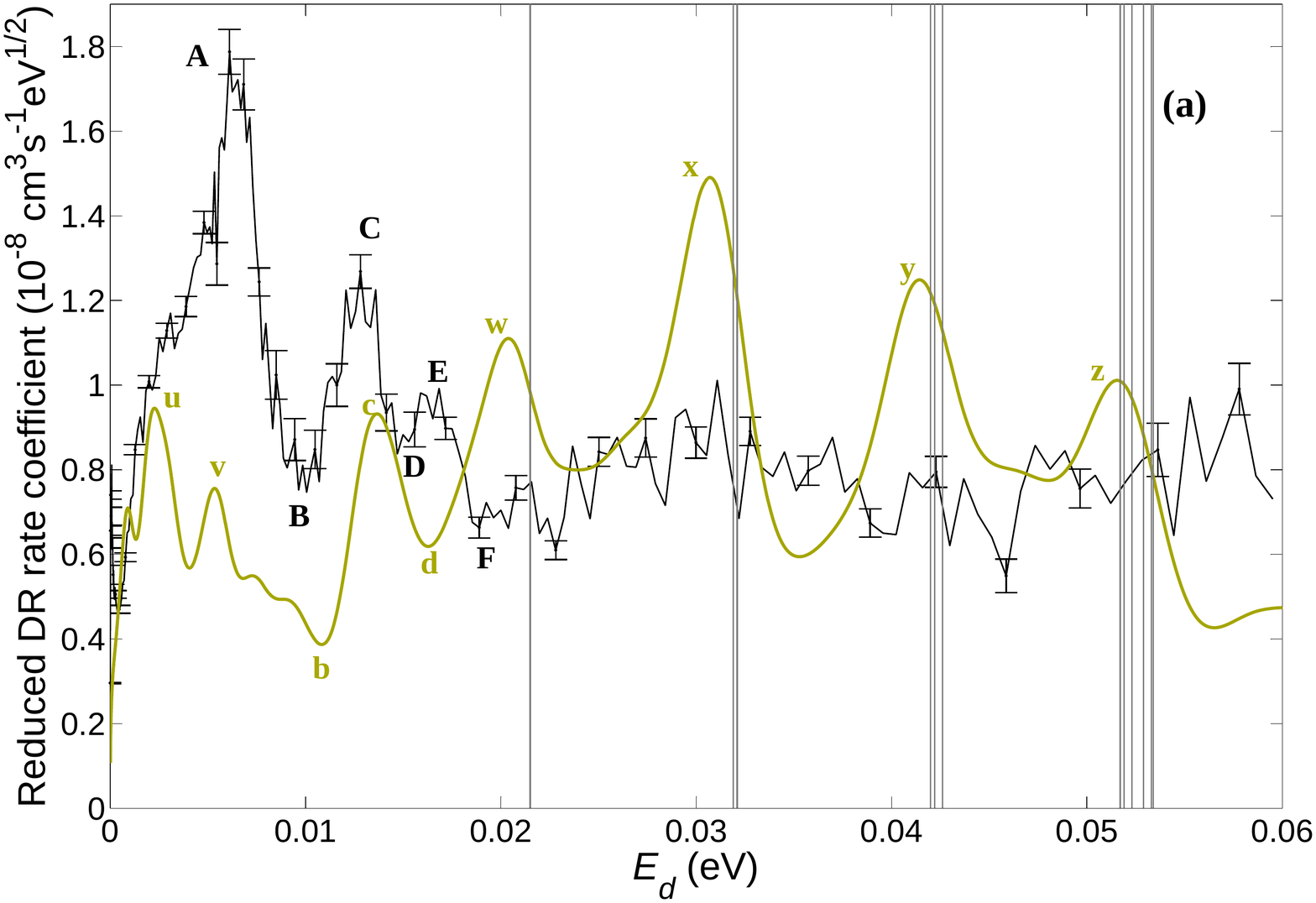}
\label{fig:first_lowE}}
\\
{\includegraphics[width=\figwidth]{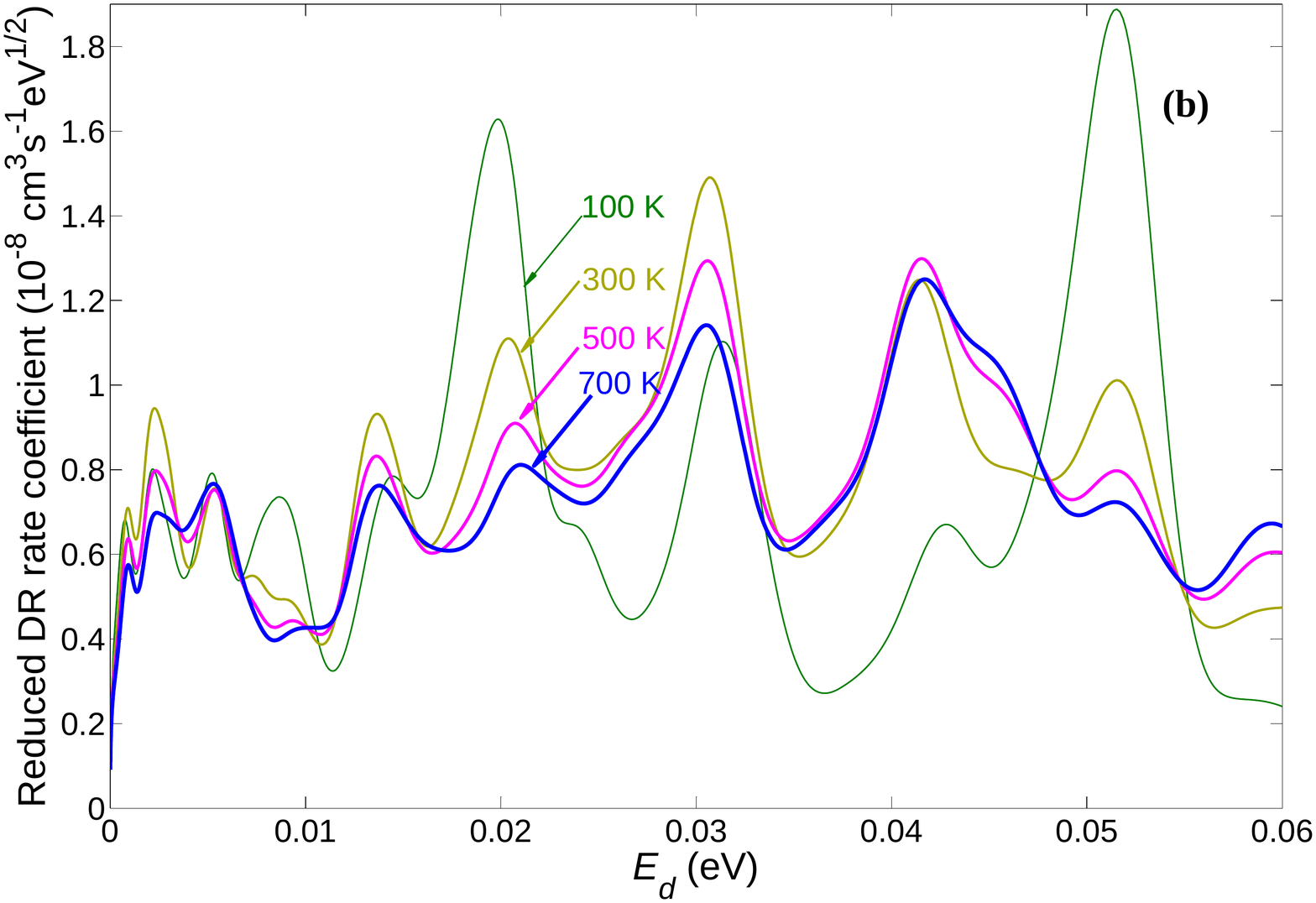}
\label{fig:second_lowE}}
\caption{\ (a) The reduced \hhh\ DR rate coefficient versus collision (detuning) energy, as measured using the beam from the 22-pole-trap (R22n), is compared with the theoretical rate coefficient calculated for a source of thermal ions at 300K, over the range from 0 to 0.06 eV. The relevant excited rotational thresholds that can produce an infinite Rydberg series of $np$-resonances just below each of them are shown as vertical lines. The numerical values of those threshold energies and their corresponding quantum numbers are shown in Table~\ref{table:thresholds}. For clarity, error bars are only shown for every 5th point. (b) Over the same energy range, theoretical spectra for the reduced DR rate coefficient are shown at four different temperatures (100K, 300K, 500K, 700K), with the line thickness increasing with temperature.}
\label{fig:ratelinlin300KlowE}
\end{figure}

\begin{figure}
\centering
{\includegraphics[width=\figwidth]{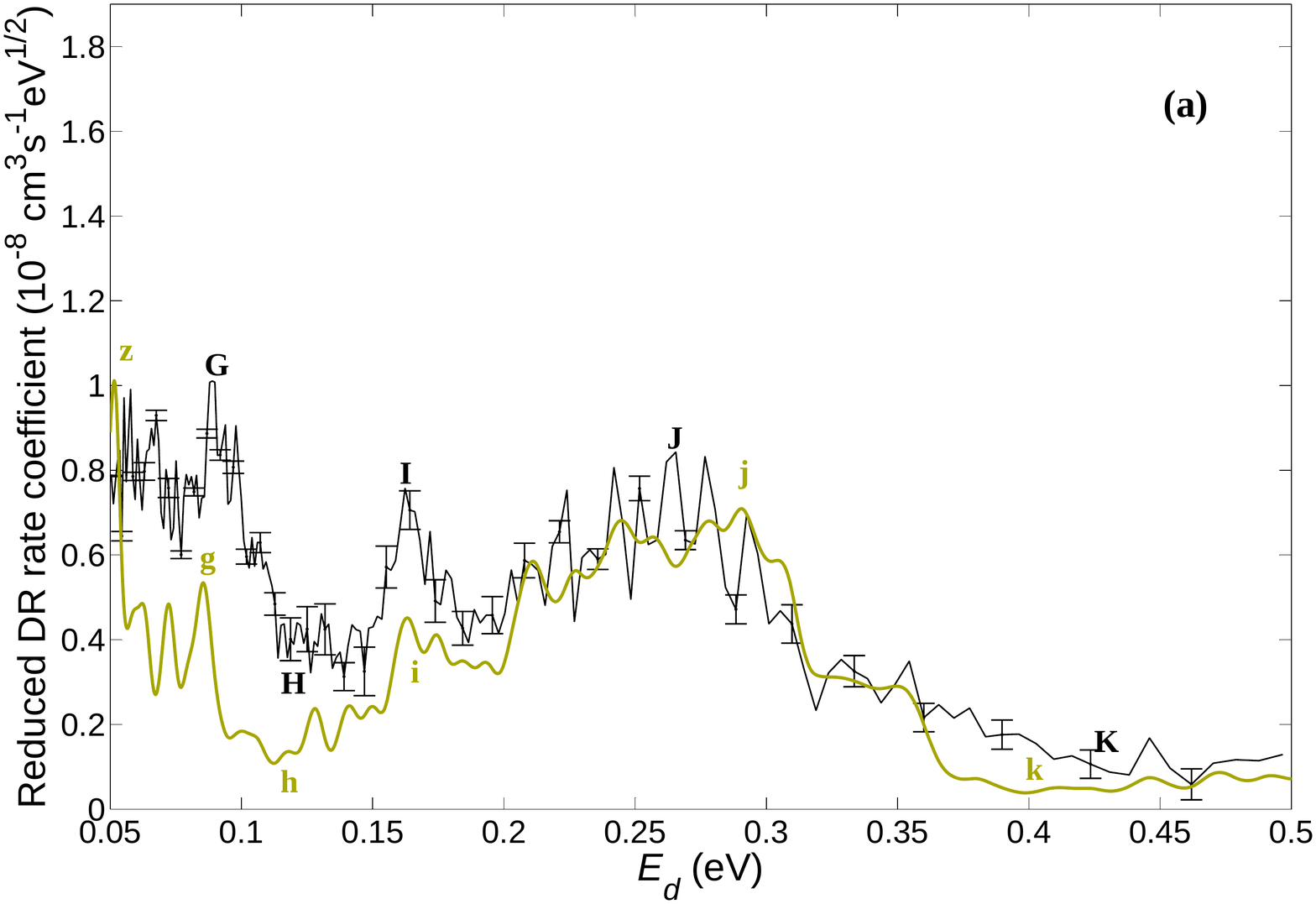}
\label{fig:first_sub}}
\\
{\includegraphics[width=\figwidth]{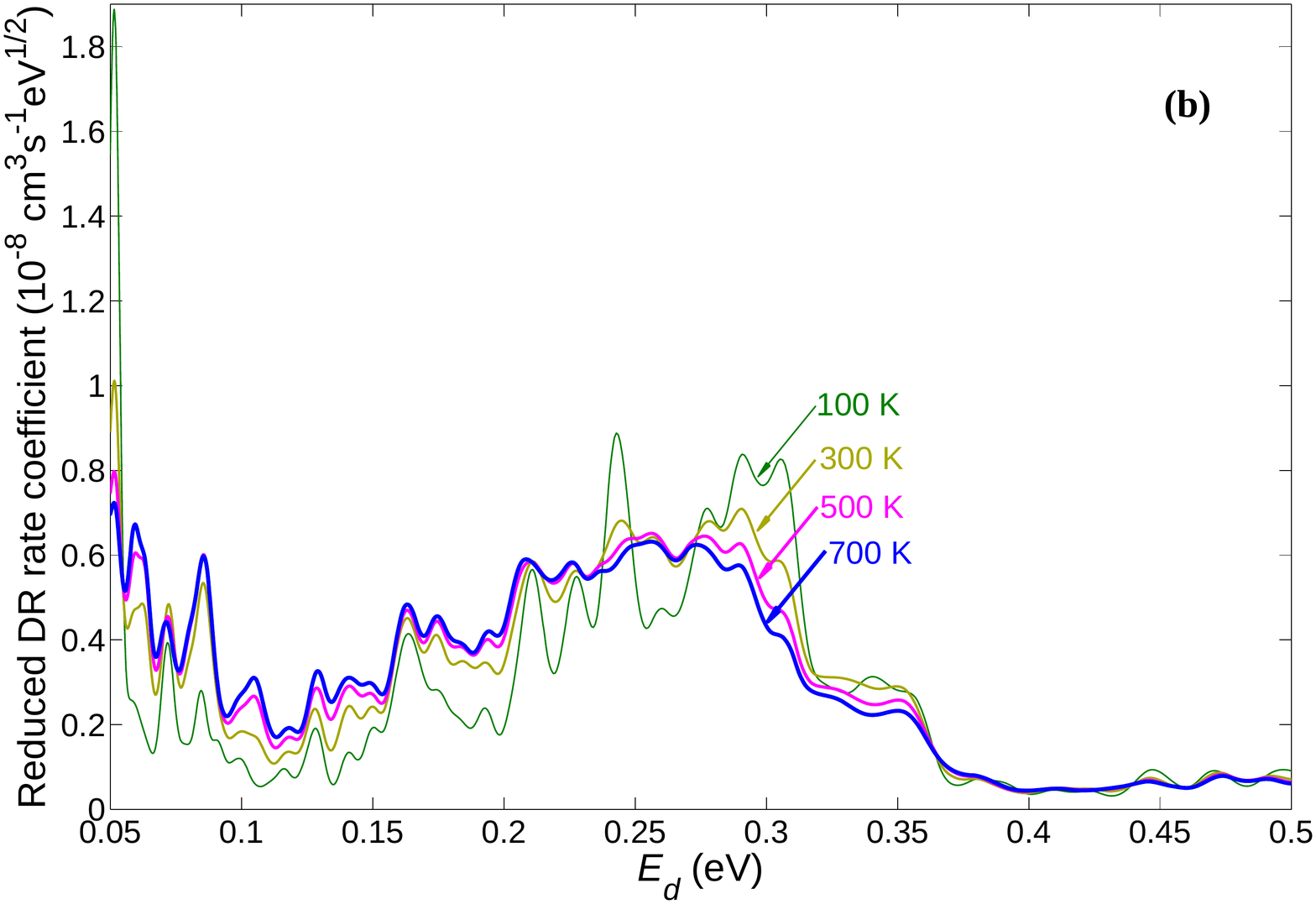}
\label{fig:second_sub}}
\caption{\ (a) The reduced \hhh\ DR rate coefficient versus collision (detuning) energy, as measured using the beam from the 22-pole-trap (R22n), is compared with the theoretical rate coefficient calculated for a sample of thermal ions at 300K, over the range from 0.05 eV to 0.5 eV. For clarity, error bars are only shown for every 5th point. (b) Over the same energy range, theoretical convolved spectra for the reduced DR rate are shown at four different temperatures (100K, 300K, 500K, 700K), with the line thickness increasing with temperature.}
\label{fig:ratelinlin300K}
\end{figure}

Figures \ref{fig:ratelinlin300KlowE}a) and \ref{fig:ratelinlin300K}a) adopt a linear-linear plot to compare the experimental reduced DR rate coefficient of rotationally cool \hhh, R22n, with theoretical reduced DR rate coefficients in greater detail, up to a collision (detuning) energy of 0.06 and 0.5 eV, respectively. The theoretical curve is calculated using an assumed \hhh\ ion temperature of 300\,K, which is not far from the estimated ion beam internal temperature as was discussed above. Parts (b) of these two figures display the theoretically expected temperature-dependent rate coefficients from 100\,K to 700\,K in the corresponding low and high energy ranges.

The following observations follow from studying Figs. \ref{fig:ratelinlin300KlowE} and \ref{fig:ratelinlin300K}:

\begin{table}
\caption[]{Ionization thresholds relevant to a $p$-wave electron colliding at low energy with the \hhh\ ion in different initial ionic energy eigenstates characterized by either ortho (A$_{2}{''}$ or A$_{2}{'}$) or para (E${''}$ or E${'}$) symmetry.} 
\label{table:thresholds}
\begin{tabular}{|c|c|c|}
\hline
Target State $(N^+,K^+)$ & Upper State & $E_{threshold}$ (meV) \\
\hline
\hline
 E${''}$ (1,1) & (2,1) & 21.5  \\
 E${''}$ (1,1) & (3,1) & 53.4  \\
 E${''}$ (2,1) & (3,1) & 31.9  \\
% E${''}$ (2,1) & (4,1) & 73.9  \\
 E${''}$ (3,1) & (4,1) & 42.0  \\
% E${''}$ (3,1) & (5,1) & 93.7  \\
 E${''}$ (4,1) & (5,1) & 51.7  \\
 E${'}$ (2,2) & (3,2) & 32.1  \\
% E${'}$ (2,2) & (4,2) & 74.3  \\
 E${'}$ (3,2) & (4,2) & 42.2  \\
% E${'}$ (3,2) & (5,2) & 94.1  \\
 E${'}$ (4,2) & (5,2) & 51.9  \\
 E${'}$ (4,4) & (5,4) & 52.9  \\
 A$_{2}{'}$ (1,0) & (3,0) & 53.3  \\
% A$_{2}{'}$ (3,0) & (5,0) & 93.5  \\
 A$_{2}{''}$ (3,3) & (4,3) & 42.6  \\
% A$_{2}{''}$ (3,3) & (5,3) & 94.9  \\
 A$_{2}{''}$ (4,3) & (5,3) & 52.3  \\
 \hline
\end{tabular}
\end{table}

{\bf (1)} Theory and experiment in the lower energy range from 0-0.06 eV exhibit clear disagreement in the shape of the reduced DR rate versus energy, as is apparent in Fig.~\ref{fig:ratelinlin300KlowE}. The largest experimental resonance (A) appears to have no theoretical counterpart, and several prominent theoretical resonances (u, v, w, x, y, and z) have no discernable experimental companions. The lone region where theory resembles experiment in this region is the range from the local minimum (B) to a maximum (C), followed by a minimum (D). The theoretical local maximum (c) is located about 1 meV higher in energy than the experimental maximum (C); it is possible that this approximate agreement is entirely fortuitous, given that this is the only feature in the lower energy range portion of the graph where the theoretical curve resembles experiment.

{\bf (2)} A working hypothesis concerning the low-energy discrepancy between theory and experiment in Fig.\ref{fig:ratelinlin300KlowE}a) is that something problematic must be occuring in the treatment or the detection of the very highest Rydberg states in theory or experiment, respectively. It is intriguing, and arguably more than a coincidence, that the 4 rightmost theoretical resonances (centered near 0.02, 0.03, 0.04, and 0.05 eV), none of which has any experimental counterpart, are lying on and slightly below a rotational threshold supporting an infinite number of high Rydberg states. It should be kept in mind that the thresholds shown as vertical lines are {\it total} electron energies, whereas the theoretical and experimental DR curves are convolved with the asymmetric electron distribution in velocity space, implying that the peaks in those curves correspond to higher total energies by approximately 0.5 meV. If we turn the width of these resonances into a range of principal quantum numbers, these regions of enhanced DR are predominantly contributed by Rydberg states having high principal quantum numbers, $n>50$. A slightly suspicious point concerning the theory is that normally one would expect that a simple one-channel Rydberg series of $p$-states converging to an excited rotational threshold of the ion should decay by autoionization rather than by dissociating \cite{CurikMolPhys:2007}. A high DR rate in this case is expected to occur only if there is a lower-$n$ perturber, e.g. attached to a $(01)^1$ state of vibrational angular momentum, straddling that threshold energy range as well. On the other hand, we cannot rule out that theory might be correct in expecting enhancements at those rotational thresholds; in this case what should be explored is the possibility that the Rydberg states are either destroyed or their angular momentum quantum numbers are changed by external fields in the storage ring.

{\bf (3)} The higher energy range of Fig.~\ref{fig:ratelinlin300K} shows a good general agreement in magnitude and overall shape between the experimental and theoretical curves, consistent with previous studies. The temperature dependence overall tends to casue a gradual increase with increasing internal temperature of the target ions, but in some localized energy ranges exceptions are evident where the lower temperature gives a higher DR rate.

{\bf (4)} The DR rate coefficient in dependence on electron energy falls off approximately as 1/$\sqrt{E_d}$ as the energy increases, over the range 0-0.3 eV, and at 0.3 eV (the first vibrational threshold as expected from the theoretical DR mechanism\cite{kokoouline03b,curik:2007,JungenPratt:2009}) the rate drops quickly from regions J and j to K and k, although somewhat more steeply in the theoretical treatments than in experiment. Both theory and experiment show a plateau region in reasonable agreement midway between features J and K.

{\bf (5)} The region of collision energies ranging from 0.05 eV to 0.5 eV exhibits one of the best overall agreements between theory and experiment in the DR of \hhh.  The shapes of both the experimental curve (from G-H-I-J-K) and the corresponding theoretical curve (ranging from g-h-i-j-k) show a level of similarity not observed before. In localized regions, however, theory is lower than experiment by 30-50\%, especially near features G, H, and I. Resonant features I and i appear to coincide closely in position, although this feature does not involve just a single isolated resonance. Instead, the theoretical calculations in this energy range indicate many resonances from a number of different symmetry classes contributing to this structure. \\

\begin{figure}
\centering
{\includegraphics[width=\figwidth]{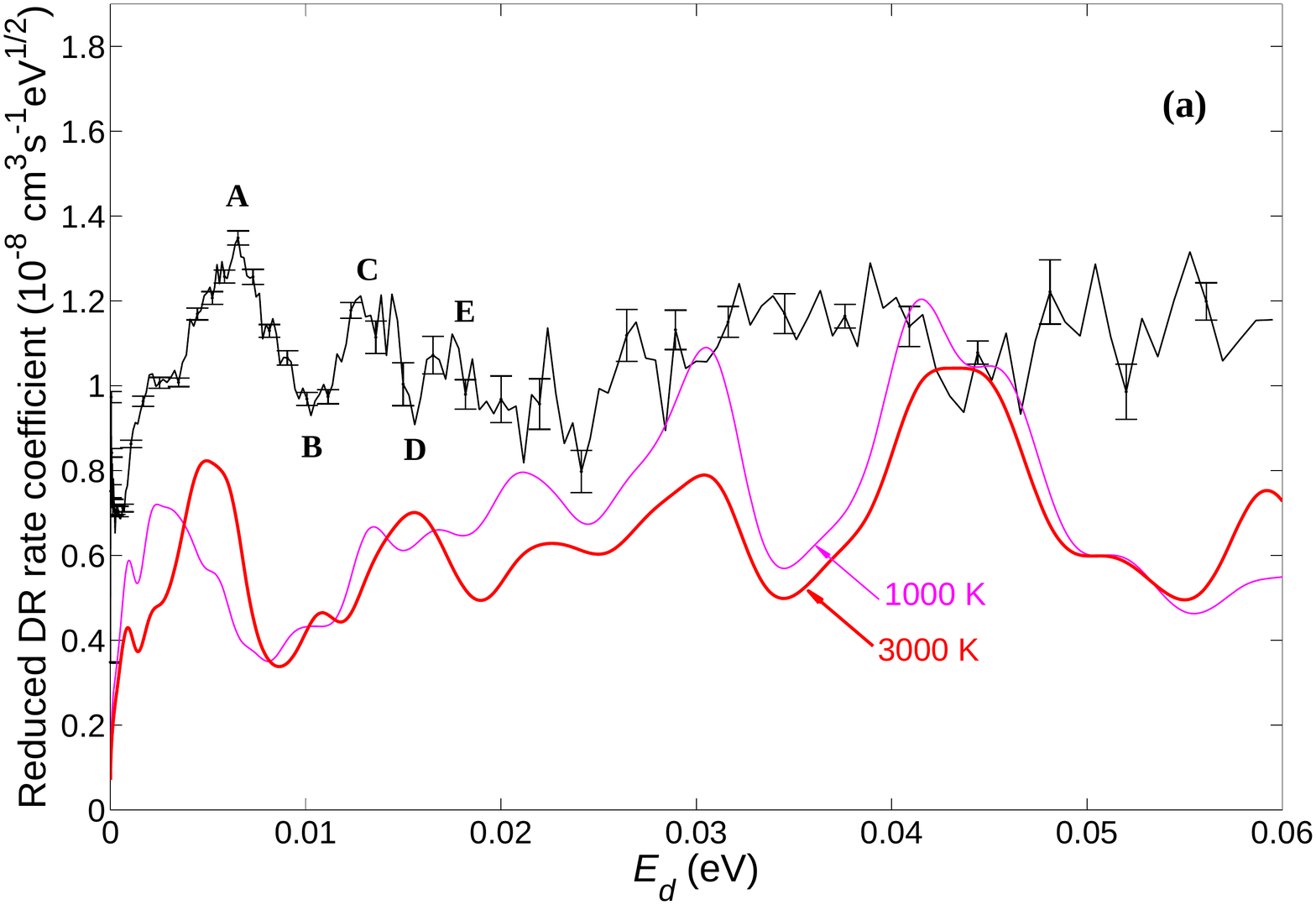}
\label{fig:lowEhot}}
\\
{\includegraphics[width=\figwidth]{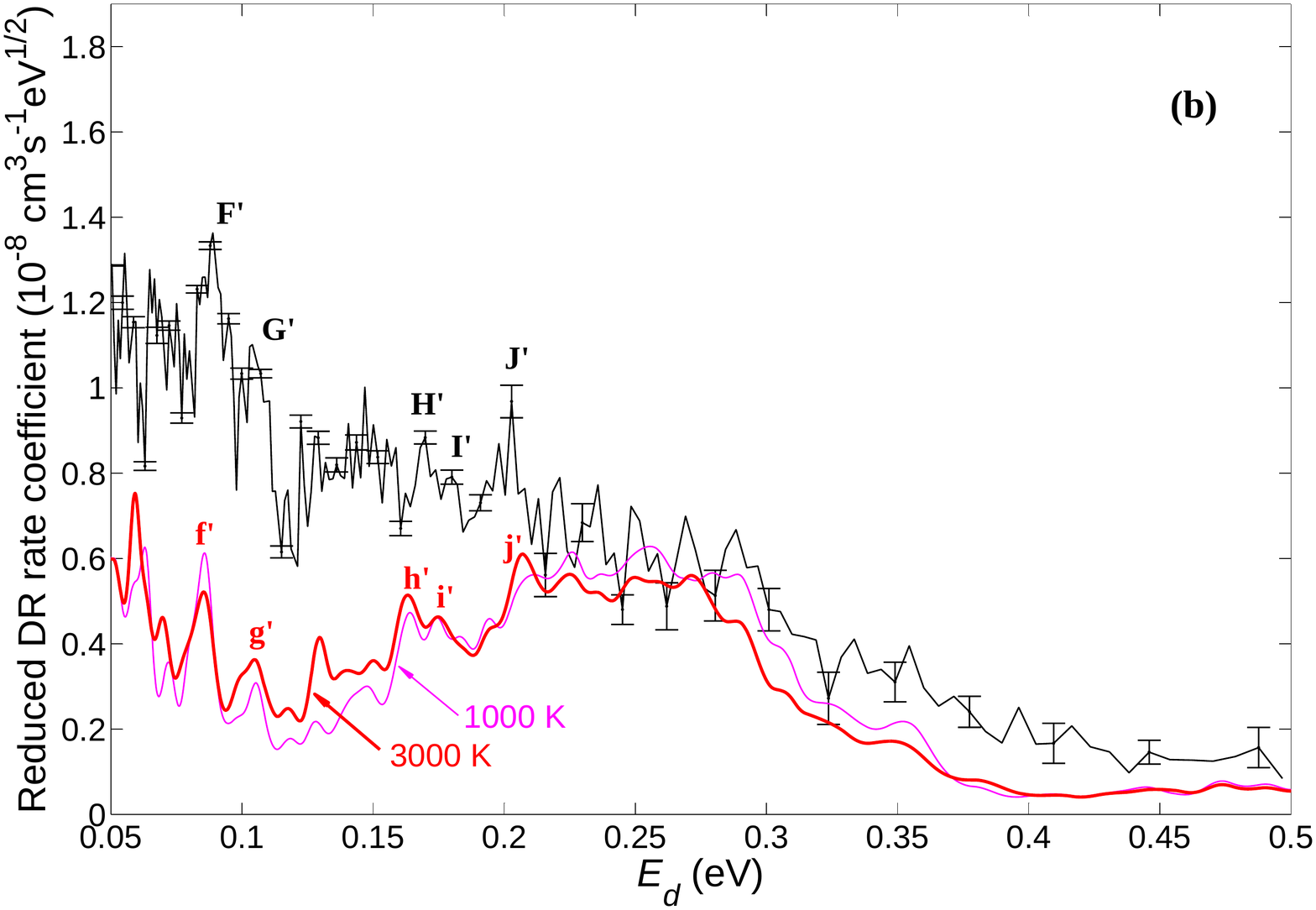}
\label{fig:highEhot}}
\caption{\ (a) The reduced \hhh\ rate coefficient versus collision (detuning) energy, as measured using the beam from the Penning source (Rpenn), is compared with the theoretical rate calculated for a source of thermal ions at 1000\,K (purple) and 3000\,K (red), over the range from 0 eV to 0.06 eV. For clarity, error bars are only shown for every 5th point. (b) The same for the energy range of 0.05 eV to 0.5 eV. The line thickness increases with temperature.}
\label{fig:ratelinlin3000K}
\end{figure}

Figures \ref{fig:ratelinlin3000K}a) and b) show a linear-linear plot of the experimental reduced DR rate coefficients of the rotationally hot \hhh, Rpenn, in comparison with theoretical reduced DR rate coefficients at 1000 and 3000\,K up to a collision (detuning) energy of 0.06 eV and 0.5 eV, respectively. The \hhh\ ion temperature of 3000\,K is close to the derived ion beam internal temperature from the imaging experiments. As the temperature matching the rate coefficient Rpenn is expected to be cooler due to the use of the electron cooler, the rate at an ion temperature of 1000\,K is shown as well. The following observations may be noted:

{\bf (6)} There is some similarity to be observed in the shape of the rate coefficient for the higher energy range as indicated by features F' to I' and f' to i'. All of these features are however a combination of resonances from different symmetries and the mere occurence of resonances together with the differing amplitudes makes any definite statement difficult. The experimental and theoretical rates are in good agreement in the energy region above J'(j'), with the theoretical rate dropping slightly more steeply at the vibrational threshold around 0.3 eV, similar to observation {\bf (4)} for the rotationally cool \hhh.

{\bf (7)} Structures A to E are observed both for rotationally cool (Fig. \ref{fig:ratelinlin300KlowE}) and hot \hhh (Fig. \ref{fig:ratelinlin3000K}b) with the amplitude of feature A increasing most with lower temperature. This could support the argument that the more pronounced structure observed in R22n and R22n-09 (see Fig.~\ref{fig:compareExps}) is due to a lower rotational ion temperature. There are several possible explanations for the consistent shape of the A-E structure at varying rotational temperature. One explanation might be that the A-E structure is a consequence of higher populated low $N^+$ levels than assumed for a $10^3$\,K Boltzmann distribution. The rotational population of the \hhh\ beam is probably not Boltzmann-distributed. Additionally, higher rate coefficients for low $N^+$ levels than for high $N^+$ levels would lead to a higher contribution from the lower levels. This could give rise to an apparent remnant cold structure in the hot \hhh\ rate coefficient, as can also be observed in Fig. \ref{fig:ExpRateslog}. Another explanation is that the seemingly invariable behaviour is fortuitous. Features A-E consist of multiple resonances and are most probably contributed by low Rydberg state perturbers ($n<9$). The contributions of low and high $N^+$ levels will therefore give rise to resonances at similar energies, producing similar though different features. This is well demonstrated in regions t' and u' where the shapes of the resonance features for 1000 and 3000\,K are similar, but where in this case the different origins of the features are revealed by a clearly visible change in position.

{\bf (8)} A larger discrepancy appears in the overall amplitude of the rate coefficient than for the low-temperature counterpart. The average value of the predicted reduced DR rate is clearly lower than the one observed, in both energy regions. Enhancement of the rotational temperature does not narrow this gap as it did in the low-temperature case. Additonal rotational states have little more effect on the total rate as can be observed by the similar average rates for 1000 and 3000\,K temperatures.  It should be kept in mind, however, that the 3000\,K theoretical rate coefficient is not fully converged with respect to the highest ionic angular momenta in the Boltzmann distribution.
 
%_________________________________________________
\section{\label{sec:disc} Conclusions}
The temperature derived by the imaging measurements for the 22-pole trap under the conditions of a rate measurement is clearly higher than had originally been expected. As suggested in the theoretical work of Ref. \cite{santos:2007}, a higher temperature of the stored ions has to be assumed for the measured rate coefficients of the rotationally cool \hhh. Although Ref. \cite{santos:2007} proposed an ion temperature of 1000\,K for the storage-ring experiments, it seems that 300\,K is sufficient to attain a similar average rate. Disagreements up to an order of magnitude are hereby reduced to a factor of roughly 2.

The experimental work reported in Ref. \cite{kreckel:2010} suggests that the extraction from the supersonic ion source could be the cause for the heating of the ion temperature (up to 950\,K). It is unclear whether ion extraction causes heating also in the 22-pole trap. The ion-beam temperature derived from previous imaging measurements, I22p-09 and I22n-09 (see Table \ref{table:ImExps}), is confirmed to be $<300$\,K using an improved simulation of the dissociation events and including an updated kinetic energy release and ion beam energy. With the statistical and systematic uncertainties amounting to roughly 135\,K, the trap temperature of $\sim15$\,K cannot be entirely excluded. Heating is, however, observed inside the ring. The $<300$\,K temperature applies to conditions where the electron cooler is used for only 0.5 s. When employing the electron cooler for the entire storage time, as was the case for IR22p, the ion temperature is raised to $\sim380$\,K, implying heating by the electron beam of the cooler. This is the opposite effect as was observed in Ref. \cite{kreckel:2010} for the supersonic ion source, where the electron cooler seems to cool the \hhh\ beam from initially $\sim950$\,K to $\sim450$\,K. The cooling of hot beams and heating of cold beams may be an indication of an equilibrium being attained around 400 K. If so, the rotationally hot \hhh\ beam of the Penning source will be colder than the $3250$\,K temperature derived by imaging (Ipenn-09) with the electron cooler on for only 0.5 s.

The \hhh\ rate coefficients measured at the storage ring TSR are all obtained with the electron cooler on cooling conditions for the entire storage time, implying $\sim380$\,K with the 22-pole trap and $\sim450$\,K with the supersonic ion source \cite{kreckel:2010}. Presently no rate-coefficient measurement with a confirmed temperature below 300\,K exists. The present work suggests that in order to obtain a rate coefficient measurement at lower ion temperatures, the electron beam of the electron cooler should be excluded from the measurement or its influence reduced by minimizing its transverse energy spread. The electron target having an electron beam with a much smaller velocity spread (1 meV) could be used for phase-space cooling the ion beam instead. Its toroidal regions are also designed to (de)merge the electron beam faster, reducing any possible temperature change from non-zero energy collisions. To eliminate possible heating from ion-source extraction, which so far can be neither confirmed nor excluded, the helium buffer-gas may be pulsed into the trap during initial cooling only, reducing the particle density upon extraction.

At the currently estimated temperature of $\sim380$\,K, 95\% of the ions are distributed over 15 rotational states up to $N^+=5$ instead of the previously reported cold rovibrational populations of predominantly $N^+=1$. These are too many rotational states to make a detailed study of the contributions of specific symmetries. DR measurements of real rotationally cold \hhh\ ion beams are necessary to advance further. The level of agreement between theory and experiment is however encouraging, especially in the energy range 0.06 eV to 0.5 eV, where a number of features that modulate the overall energy-dependent DR rate coefficient appear to be mirrored in theory and experiment. This comparison is the most detailed one carried out to date. Nearly all of the major discrepancies at collision energies below 0.06 eV arise in a narrow range approximately 4 meV below ionization thresholds supporting very high Rydberg states. This points to the need of combined efforts by theory and experiment in the range of very high Rydberg principal quantum numbers $n>50$ in order to track down the origin of this major disagreement.

The present analysis stresses the need for continuing efforts to develop an {\it in-situ} method for determining and if possible controlling the rotational state distribution of the ions in the storage ring. Preliminary work on an {\it in-situ} probing procedure is being performed \cite{petrignani:2010}. Additionally, the predicted 10:1 higher rate coefficient for the para-(1,1) than for the ortho-(1,0) state can be studied in more detail when low temperatures are attained. Further improvements to the theory which are under consideration elsewhere are (i) the elimination of the currently-implemented rigid-rotator approximation for the \hhh\ rotational degrees of freedom; (ii) inclusion of $l-mixing$ effects due to the non-spherically-symmetric interaction experienced by an electron that collides with the ion; and (iii) continued exploration of external field effects on Rydberg-mediated DR.

%_________________________________________________
\begin{acknowledgments}
This work was partially supported by the National Science Foundation, by the DOE Office of Science, and by an allocation of NERSC supercomputing resources. Support from the Max-Planck Society is acknowledged. VK also thanks the ``R\'eseau Th\'ematique de Recherche Avanc\'ee'' {\it Triangle de la Physique} for partial support. HB acknowledges partial support from the German Israeli Foundation for Scientific Research and Development (G.I.F.) under Grant I-900-231.7/2005 and by the European Project ITS LEIF (HRPI-CT-2005-026015). ON was supported in part by the NSF Astronomy and Astrophysics Grant No. AST-080743. We are grateful to the TSR accelerator group for their support during the experiments.
\end{acknowledgments}

%\bibliography{\bibpath H3plus}
\bibliography{H3plus}

\end{document}